\documentclass[twocolumn,aps,prc,superscriptaddress,showpacs,floatfix]{revtex4}

\usepackage{epsfig}
\usepackage{amsmath}
\usepackage{multirow}
\usepackage{graphicx}
\usepackage{amsmath}

\usepackage[normalem]{ulem}  % \sout{old text} for strikeout
\usepackage[dvips]{color} % For blue in-text comments and additions

\renewcommand\sout{\bgroup \color{red} \ULdepth=-.5ex \ULset}

\newcommand{\Ex}[2]{\ifmmode{#1\times10^{#2}}\else{$#1\times10^{#2}$}\fi}
\begin{document}
\title{Diquarks: a QCD sum rule perspective}

\author{Kyungil Kim}\affiliation{Institute of Physics and Applied Physics, Yonsei
University, Seoul 120-749, Korea}
\author{Daisuke Jido}\affiliation{Yukawa Institute for Theoretical Physics, Kyoto University, Kyoto 606-8502, Japan}
\author{Su Houng Lee}\affiliation{Institute of Physics and Applied Physics, Yonsei University, Seoul 120-749, Korea}\affiliation{Yukawa Institute for Theoretical Physics, Kyoto University, Kyoto 606-8502, Japan}

\date{\today}
\begin{abstract}
We propose a phenomenological QCD sum rule with an explicit diquark
field to investigate the essential ingredients inside the hadrons. By introducing the mass $m_\phi$  and the condensate $\langle \phi^2
\rangle$ for the diquark field as parameters in the model, we find
that the sum rule works well for $\Lambda$, $\Lambda_c$ and
$\Lambda_{b}$.  This implies that these $\Lambda$ baryons can be
represented by a diquark and a quark configuration. We also find
that there is a duality relation among the parameters ($m_\phi,
\langle \phi^2 \rangle$) for which the sum rule is equally good.
In the limit when $\langle \phi^2 \rangle =0$, we find $m_\phi=400$
MeV, which can be thought of as the constituent diquark mass.
%we conclude that there is essentially a single parameter in the diquark sector controlling dynamics.
%the $\Lambda$ masses are essentially controlled by a single quantity, such as the constituent diquark mass,
%which is obtained as $m_\phi=410$ MeV in the limit when $\langle \phi^2 \rangle =0$.
%when $\langle \phi^2 \rangle$ decreases, $m_\phi$ has to increase to recover the stability. Finally, in the limit when $\langle \phi^2 \rangle =0$ we find $m_\phi=410$ MeV, which can be thought as the constituent diquark mass.
\end{abstract}

\maketitle

\section{Introduction}

Quests for fundamental correlations in strongly interacting systems
can be a clue to understanding the structure of the physical
systems. For instance, the Cooper pair, which is a correlated
electron pair, is a key concept to understand many-electron systems
and leads to  superconductivity after it condenses with other pairs.
Also in nuclear physics, two-nucleon pair correlation can be an
important degree of freedom to describe the nucleus, such as in the
interacting boson model. In hadron physics, two-quark pair
correlations, or diquarks, were introduced in the early days of
quark model to explain systematics in baryon
spectroscopy\cite{Ida:1966ev,Lichtenberg:1967zz}.  Over the years,
experimental findings in the structure function,
fragmentation function and weak decay processes pointed to the
existence of a strong diquark correlation, especially in a definite
spin-flavor configuration, which is the so-called good diquark
channel~\cite{Jaffe:2004ph}.

In two-color QCD with massless quarks, interquark interactions have
exactly the  same strength as $\bar q$-$q$ correlations due to the
Pauli-G\"ursey symmetry. This symmetry is explicitly broken in
three-color QCD, but the remnant is seen in the good diquark
channel, which has a resemblance but less strong attraction to its
counterpart of $\bar qq$. The strong diquark correlation in the
good channel can be derived from one gluon exchange potential or
through the instanton-induced interaction, and is  found in the lattice
calculation  in the Landau gauge~\cite{Hess98} and  Coulomb
gauge~\cite{Babich}.  It is also found in the configuration with an
additional  color source to make the whole configuration gauge
invariant~\cite{Orginos,Alexandrou}. However, in most hadron
configurations the strong diquark correlation expressible as an
elementary constituent has not been so evident. The reason why
diquarks cannot be treated as an elementary constituent can be
traced back to the original work by Jaffe on the tetraquark picture
for the scalar nonet~\cite{Jaffe76-1}.  There, the wave function for
$\sigma$ is found to be dominantly composed of the $\pi$-$\pi$ state
in the meson-meson basis, which translates into a small
diquark-antidiquark component in their respective good channels.
This means that when an antiquark is present, the diquark would
break and form a quark-antiquark pair, which is  energetically more
favorable  than a diquark.  This phenomenon is related to the reason we
have quark condensate over diquark-antidiquark condensate in the QCD
vacuum.

Nonetheless, there are certain configurations where the good
diquark  correlation is the strongest and its role is important.  The
color superconducting phase appearing at high density is an
example~\cite{Alford:1997zt,Rapp:1997zu}.  There the strong
correlation in the good diquark channel is responsible for the
Cooper pairing.  At the hadron level, the simplest configuration of
this sort is the baryon composed of a diquark $(ud)$ and a
spectator quark ($h$); namely
$\Lambda_h$, where $h$ can be $s,c$ or $b$ quarks.
The reason that a diquark correlation remains strong can be understood from the color-spin interaction, which is known to be inversely proportional to the constituent mass of the quarks. Inside the $\Lambda$, the color-spin interaction between $u-h$ quarks cancels that between $d-h$ quarks when the diquark ($ud$) is in the I=0 channel.  Moreover, the color-spin interactions with the heavy quark itself become smaller as $h$ becomes heavy
\cite{Rosner}.  Hence the diquark will retain its strong correlation with small $uh$ or $dh$ correlations.  Other configurations,
where diquark correlations are expected to remain important are hadrons with large angular momentum~\cite{Selem:2006nd} and exotic configurations with heavy quarks~\cite{Lipkin87,Lee07,Lee09}.

In this paper, we will show that, inside the $\Lambda_h$, diquark
correlation is indeed strong so that the diquark can be treated as
an elementary constituent.  We further show that one can extract
well-defined parameters for the {\it diquark}, from which one can
also define the constituent diquark mass.

\section{Formulation}

To carry out our task, we propose a framework for describing the
$\Lambda_h$ based on the QCD sum rule
technique~\cite{Shifman:1978bx,Shifman:1978by}. In the usual
study of baryon spectra in the QCD sum rule, one calculates a
correlation function of baryonic currents composed of the quark
fields using the operator product
expansion~\cite{Ioffe:1982ce}. If the correlation of a pair
of quarks is strong, the pair can be regarded effectively as 
elementary degrees of freedom. Thus, we introduce an explicit
scalar-diquark field $\phi^\dagger_a$ with the color charge $a$ in
$\bar{3}$ and calculate the correlation function of the
$\Lambda_{h}$ current given by $J_\Lambda=\phi^\dagger_a h^a$ with
$h$ being the heavy quark field:
\begin{eqnarray}
\Pi(q) & = & i \int d^4 x e^{iqx} \langle T[J_\Lambda(x), \bar{J}_\Lambda(0) ] \rangle  \nonumber \\
& =& \hat q \Pi^e(q^2) + \Pi^o(q^2), \label{def}
\end{eqnarray}
where $\hat q \equiv q^{\mu}\gamma_{\mu}$.

The dynamics of the diquark field obeys the color SU(3) gauge theory:
\begin{eqnarray}
L_\phi=\phi^\dagger [D^2+m^2_\phi] \phi,  \label{eq:diqQCD}
\end{eqnarray}
where the covariant derivative for the diquark is given by
$D_\mu=\partial_\mu -ig t^a A_\mu^a$ and $m_{\phi}$ is the diquark
mass. We incorporate Eq.~(\ref{eq:diqQCD}) into the QCD Lagrangian.
As we emphasized before, we will be using this Lagrangian only in
configurations where the diquark correlations remain strong.
One
should note that the high energy limit of the correlation in
Eq.~(\ref{def}) does not have a corresponding high-energy limit in
real QCD, as the diquark field is treated here as an elementary
field, which is not the case in reality. Nevertheless, one can use
this framework to check the  validity of the description of
$\Lambda_h$ by confirming that the current $J_{\Lambda}$ couples
strongly to the physical $\Lambda_h$ at low energy and that the
correlation function reproduces the correct mass of $\Lambda_h$.

The operator product expansion (OPE) makes it possible to write down
the correlation function in terms of vacuum expectation values of
the elementary fields stemming from nonperturbative dynamics. The
OPE of both the even and odd part of the correlation function
(\ref{def}) can be calculated using the fixed-point gauge.  We
include the diquark condensate $\langle \phi^\dagger \phi \rangle
\equiv \langle \phi^{2} \rangle $ and perform the calculation to
leading order in $\alpha_s$ up to terms of  condensates having
dimension 6. Note that since the diquark field has dimension 1 in
the $\Lambda_{h}$ current  $J_{\Lambda}= \phi^{\dagger} h$, the
correlation function (\ref{def}) has mass dimension 1, which is
smaller than the usual baryonic correlators.

The perturbative part is as follows:
\begin{subequations}
\label{eq:pert}
\begin{eqnarray}
{\rm Im}\Pi^e_{\rm p} & = & \frac{3\pi}{(4\pi)^2} \frac{\tilde{q}^2 }{2q^2} \frac{\bar{q}^2}{q^2} u \\
{\rm Im} \Pi^o_{\rm p} & = & m_h \frac{3\pi}{(4 \pi)^2}  \frac{\bar{q}^2}{q^2} u,
\end{eqnarray}
\end{subequations}
where $\bar{q}^2 \equiv q^2-(m_h-m_d)^2$, $\tilde{q}^2 \equiv
q^2+m_h^2-m_\phi^2$ and $u^2 \equiv 1-4m_h m_\phi/\bar{q}^2$. It is
worth noting that, although we call Eq.~(\ref{eq:pert}) the
perturbative part that is obtained by the perturbative calculation of
our effective model, since we introduce the diquark as an elementary
field, these terms contain highly non-perturbative interactions in
the sense of original QCD.

The condensate parts are given as
\begin{eqnarray}
%\lefteqn{\Pi^e =} &&  \nonumber \\
%&&
\Pi^e_{\rm c}  &=&
C_\phi^e
 \langle \phi^2 \rangle
+ C_h^e \langle \bar{h}h \rangle
+ \bigg[  \frac{3m_\phi^2 }{6 \cdot 24}
%\int_0^1 dx  \frac{x^4}{(\Delta^2)^3}
F^{40}_{3}
-\frac{C_\phi^e}{96m_\phi^2}  \bigg] \langle \frac{ \alpha}{ \pi} G^2 \rangle
\nonumber \\
&&
+ \bigg[  -\frac{3 m_h^2}{12^2}
%\int_0^1 dx  \frac{-x(1-x)^3}{(\Delta^2)^3}
F_{3}^{13}
+\frac{C_h^e}{12m_h} \bigg] \langle\frac{ \alpha}{ \pi} G^2 \rangle
 - \frac{m_\phi^2}{8D_{\phi}^3}
\langle  g\bar h \sigma  \cdot G h  \rangle
\nonumber \\
&&
- \frac{1}{9}
\bigg( \frac{2}{D_{h}D_{\phi}} +\frac{1}{D_{\phi}^2} +\frac{4(q^2+m_h^2)}{D_{h}^2q^2} \bigg) \langle g^2 \phi^2 \bar{h}h \rangle, \label{pie}
\end{eqnarray}
where $D_{\phi} \equiv q^{2}- m_{\phi}^{2}$, $D_{h} \equiv q^{2}- m_{h}^{2}$,
$F_{\ell}^{mn} \equiv \int_{0}^{1} dx (x^{m}(1-x)^{n})/(\Delta^{2})^{\ell}$
with $\Delta^2 \equiv x(x-1)q^2+(1-x)m_h^2+xm_\phi^2$, and
\begin{equation*}
C_\phi^e   =  -\frac{1}{4m_\phi^2 q^4} \bigg[ q^4-(m_\phi^2-m_h^2)^2+2m_\phi^2 q^2-\tilde{q}^2\bar{q}^2u \bigg] ,
\end{equation*}
\begin{equation*}
C_h^e  =   -\frac{m_h}{(4m_h^2q^2)^2}  \bigg(2m_h^2q^2-(\tilde{q}^2)^2 + \tilde{q}^2 \sqrt{(\tilde{q}^2)^2-4m_h^2q^2} \bigg) ,
%F_{\ell}^{mn} &=& \int_{0}^{1} dx \frac{x^{m}(1-x)^{n}}{(\Delta^{2})^{\ell}}
\end{equation*}
for the even part, and
\begin{eqnarray}
\lefteqn{\Pi^{o}_{\rm c} =} && \nonumber \\ &&
%\Pi^o & = &
C_\phi^o \langle \phi^2 \rangle
+ C_h^o
 \langle \bar{h}h \rangle
+ \bigg[  \frac{3m_h m_\phi^2}{6 \cdot 24}
%\int_0^1 dx \frac{x^3}{(\Delta^2)^3}
F^{30}_{3}
-\frac{C_\phi^o}{96m_\phi^2} \bigg] \langle \frac{ \alpha}{ \pi} G^2 \rangle
\nonumber \\
&&
+\bigg[ \frac{3 m_h }{12^2}
%\int_0^1 dx \frac{-(1-x)^3xm_h^2-x^2(1-x)^2m_\phi^2}{(\Delta^2)^3}+\frac{(1-x)^2}{(\Delta^2)^2}
(-m_{h}^{2} F^{13}_{3}-m_{\phi}^{2}F^{22}_{3} +F^{02}_{2})
+\frac{C_h^o }{12m_h}\bigg] \langle \frac{ \alpha}{ \pi} G^2 \rangle
\nonumber \\
&&
+ \frac{1}{3^3  \cdot 2}  \frac{1}{D_{\phi}^3} \langle  g^2(\bar{h}h)^2 \rangle
%\nonumber \\
%&&
 -\frac{2m_{h}}{9D_{h}q^{2}}
\bigg(  \frac{1}{D_{\phi}} +\frac{4}{D_{h}} \bigg) \langle g^2 \phi^2 \bar{h}h \rangle
% -\frac{1}{9}
%\bigg(  \frac{2m_h}{D_{h} D_{\phi}q^2} +\frac{8m_h}{D_{h}^2q^2} \bigg) \langle g^2 \phi^2 \bar{h}h \rangle
\nonumber \\
&&+ \frac{m_\phi^2}{D_{\phi}^4} \bigg(
\frac{1}{3^3 2^4}
\langle g^{2} (\bar hh)^{2} \rangle
%g_s^2  \langle \bar{h} h \rangle^2
-\frac{m_h}{2^6}  g \langle  h \sigma \cdot G h  \rangle\bigg), \label{pio}
\end{eqnarray}
where
\begin{eqnarray}
C_\phi^o  & =& -\frac{m_h}{2m_\phi^2 q^2} \bigg( q^2+m_\phi^2-m_h^2 -\bar{q}^2u \bigg)
\nonumber \\
C_h^o & = &  -\frac{1}{8m_h^2q^2} \bigg(\sqrt{(\tilde{q}^2)^2-4m_h^2q^2} -\tilde{q}^2 \bigg), \nonumber
\end{eqnarray}
for the odd part. In each part, the first contribution to the gluon
condensate comes from gluon operators emanating from the diquark and
the second contribution comes from the quark propagator. The factors of the gluon
condensate proportional to $C_\phi$ $(C_h)$ subtract the zero
momentum part of the diquark (quark) propagator contributing to the
gluon condensate, which should be taken into account through the
diquark (quark) condensate.   The  parts to be subtracted can be
obtained by the contributions from the diquark (quark) condensate
after expressing the condensate in terms of the gluon
condensate\cite{Reinders:1984sr}:
\begin{eqnarray}
\langle \bar{h} h \rangle & = & -\frac{1}{12m_h} \langle \frac{ \alpha}{ \pi} G^2 \rangle \label{hh} \\
\langle \phi^\dagger \phi \rangle & = & \frac{1}{96m_\phi^2} \langle \frac{ \alpha}{ \pi} G^2 \rangle . \label{pp}
\end{eqnarray}
%%%
In fact, these equations are obtained as the leading terms of the
heavy mass expansions. Therefore, for the $\Lambda_c, \Lambda_b$ sum
rule, we can use Eq.~(\ref{hh}) in Eq.~(\ref{pie}) and in
Eq.~(\ref{pio}) and obtain their sum rule without heavy quark
condensate.

Both the even and odd part of the correlation function satisfies the
following Borel transformed dispersion relation: 
\begin{eqnarray}
{\rm B.T.} \Pi(q^{2}) = \frac{1}{M^2} \int_{(m_\phi+m_h)^2}^\infty ds \frac{1}{\pi} {\rm Im} \Pi(s) e^{-s/M^2},
\end{eqnarray}
where B.T. stands for Borel transformation and $M$ is the Borel
mass. After the Borel transformation, the correlation function is
expressed as an expansion in $1/M^{2}$ such that higher dimension
operators have more powers of $1/M^{2}$. The Borel transformation of the
fifth term of Eq.~(\ref{pie}) and the last term of Eq.~(\ref{pio})
proportional to $m_{\phi}^{2}$ give contributions with a factor of
$\exp[{-m_{\phi}^{2}}/M^{2}]/M^{8}$, which are of a much higher order
in the $1/M^{2}$ expansion. Thus we neglect these terms in our
calculation.

We assume that the $\Lambda_h$ contribution to the correlation
function in Eq.~(\ref{def}) is given by
\begin{equation}
\Pi(q) = \frac{f(\hat q + m_{\Lambda})}{q^2-m_\Lambda^2+i\epsilon} +
\theta(q^{2}-s_{0}) ( \hat q\Pi_{\rm OPE}^{e} + \Pi_{\rm OPE}^{o}) ,
\label{pole}
\end{equation}
where $f$ is a normalization of the $\Lambda_{h}$ propagator,
$\Pi_{\rm OPE}^{i=e,o} \equiv  \Pi^{i}_{\rm p} + \Pi^{i}_{\rm c}$ is
the correlation function calculated by OPE and given in
Eqs.(\ref{eq:pert}), (\ref{pie}) and (\ref{pio}), and $s_{0}$ is the
threshold parameter that represents the energy scale where the
quark-hadron duality ansatz works in the QCD sum rule.  Our theory will only comprise a limited phase space of the whole QCD.  However, once we introduce the diquark field as an effective degree of freedom, the spectral density at high energy can be calculated perturbatively as asymptotic freedom remains valid also in scalar QCD.

In principle, the sum rules for the $\Lambda_{h}$ mass can be obtained by the
following:
%\begin{eqnarray}
%m_\Lambda= { {\bf B.T.} \bigg( \Pi^o-\Pi^o(continuum) \bigg) \over
%{\bf B.T.} \bigg( \Pi^e-\Pi^e(continuum) \bigg) } \label{borel1}
%\end{eqnarray}
%or
%\begin{eqnarray}
%m_\Lambda^2= { M^2 \frac{d}{dM^2} M^2 {\bf B.T.} \bigg( \Pi^i-\Pi^i(continuum) \bigg) \over
%{\bf B.T.} \bigg( \Pi^i-\Pi^i(continuum) \bigg) }, \label{borel2}
%\end{eqnarray}
\begin{equation} m_\Lambda= { {\bf B.T.} \big[ \theta(s_{0}-s)
\Pi^o(s)\big] \over {\bf B.T.} \big[ \theta(s_{0}-s) \Pi^e(s)\big] }
\label{borel1}
\end{equation}
or
\begin{equation}
m_\Lambda^2= { M^2 \frac{d}{dM^2} M^2 {\bf B.T.} \big[ \theta(s_{0}-s) \Pi^i(s)\big]\over
{\bf B.T.} \big[ \theta(s_{0}-s) \Pi^i(s)\big]}, \label{borel2}
\end{equation}
where $i=e$ or $o$.  It should be noted however, that $\Pi^o$ is
proportional to $m_h$ except for the term proportional to $\langle
\bar{h} h\rangle$, and hence is dominated by a single term when $m_h$ is
small.  Therefore, $\Pi^e$ should not be reliable for $\Lambda$. In
such a case, one should use Eq.~(\ref{borel2}) with $i=e$ to
calculate the mass.  Moreover, as a natural extension of the $\Lambda$ sum rule, we will use  Eq.~(\ref{borel2}) with $i=e$ also for  $\Lambda_c$ and $\Lambda_b$.

We use conventional values of the quark and gluon condensates and
the quark mass, which were determined in other QCD sum rules. For
the case of $h=s$, we use $\langle \frac{ \alpha}{ \pi} G^2 \rangle
=  (0.33\, {\rm GeV})^4$, $\langle \bar{s} s \rangle  =  0.8\,
\langle \bar{q} q \rangle =   -0.8\, (0.23{\rm GeV})^3$, 
%$\langle \bar s g \sigma \cdot G s \rangle = m_{0}^{2} \langle \bar s s \rangle$ with $m_{0}=***$ GeV
and
$m_s  =  0.12$ GeV~\cite{QCDparam}. For the higher dimensional condensates, we adopt the vacuum saturation hypothesis.   That is $ \langle g^2 (\bar{h} h)^2 \rangle = \kappa g^2 \langle \bar{h} h \rangle^2$ \cite{Hatsuda:1992bv,Klingl:1997kf}, , with $g=\kappa=1$.  We will also show that the result with different values of $\kappa$ gives  almost the same result as the sum rule has only a weak dependence on these operators.
For the sum rules of $\Lambda_{c}$ and $\Lambda_{b}$, we use $m_{c}=1.47$ GeV and $m_{b}=4.6$ GeV~\cite{cmass}, and the values of the condensates obtained by Eq.~(\ref{hh}).

%\begin{subequations}
%\begin{eqnarray}
%\langle \frac{ \alpha}{ \pi} G^2 \rangle & = & (0.33 {\rm GeV})^4 \\
%\langle \bar{s} s \rangle & = & 0.8 \langle \bar{q} q \rangle =   -0.8 (0.23{\rm GeV})^3 \\
%m_s & = & 0.12 {\rm GeV}.
%\end{eqnarray}
%\end{subequations}

The diquark quantities $m_{\phi}$ and $\langle \phi^{\dagger} \phi
\rangle$ are not determined yet and can be our model parameters to
be determined  within our framework. If the diquark description of
$\Lambda_{h}$ is valid, the sum rule should  work well and reproduce
the observed $\Lambda_{h}$ mass with reasonable values of the
diquark parameters.  To parametrize the variation of the diquark
condensate, we introduce a scaling factor $\lambda$ with respect to
its value obtained in the heavy diquark mass limit given in
Eq.~(\ref{pp}) with $m_\phi=0.2$ GeV:
\begin{equation}
\langle \phi^\dagger \phi \rangle  =  \lambda \frac{1}{96 (0.2 {\rm GeV} )^2}
\langle \frac{ \alpha}{ \pi} G^2 \rangle
\end{equation}

\begin{figure}
\includegraphics[width=4.25cm,height=4.7cm]{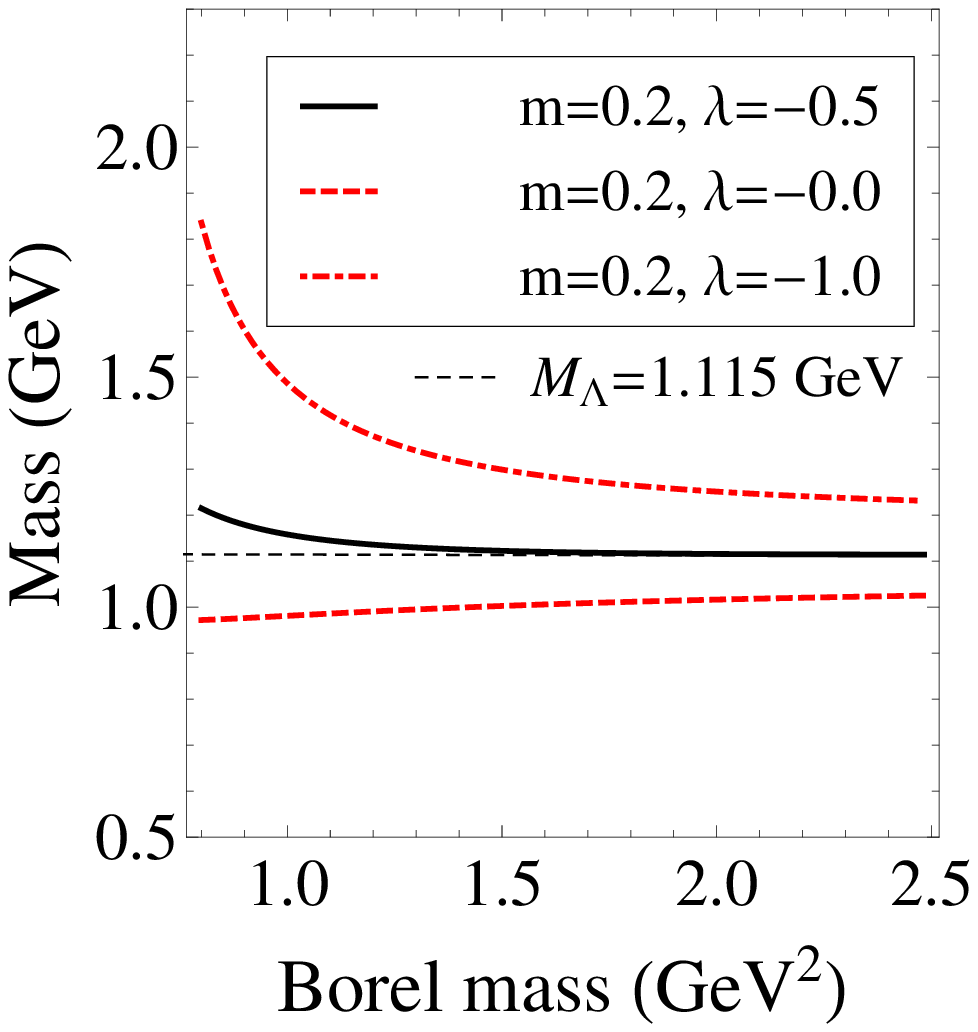}
\includegraphics[width=4.25cm,height=4.7cm]{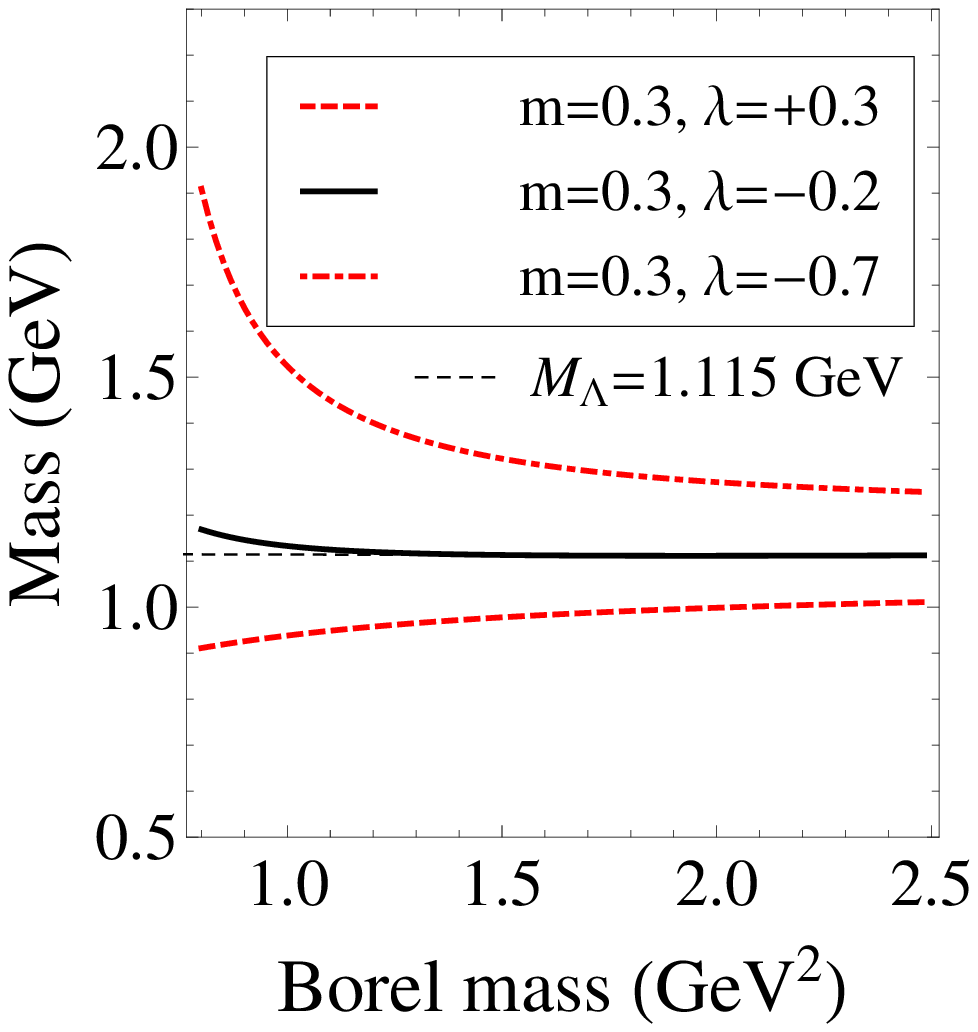}
%\Psfig{8cm}{Lambda4-0119.eps}
%~\Psfig{8.5cm}{LHC-mass.eps}
\caption{(Color online) Borel curves for $\Lambda$ with several sets of paramters.}
\label{fig-lambda}
\end{figure}

\section{Results}

\subsection{$\Lambda$}

The plots in Fig.~\ref{fig-lambda} show the Borel sum rules for the
$\Lambda$ mass obtained by using  $\Pi^e$ in Eq.~(\ref{borel2}) for
diquark masses of 0.2 and 0.3 GeV for the upper and lower figure
respectively with several condensate values.  One finds that for the
upper curve, the $\Lambda$ mass is well reproduced with a stable
Borel curve for the diquark condensate value of $\lambda=-0.5$,
which corresponds to $\langle \phi^{2}\rangle = -(0.04\, {\rm
GeV})^{2}$. The fact that good Borel stability has been obtained
implies that the correlation function of $J_{\Lambda}$ with the
diquark field strongly couples to the low-lying $\Lambda$ baryon and
thus the description of $\Lambda$ in terms of the diquark degrees of
freedom works well.

\begin{figure}
\includegraphics[width=4.25cm,height=4.7cm]{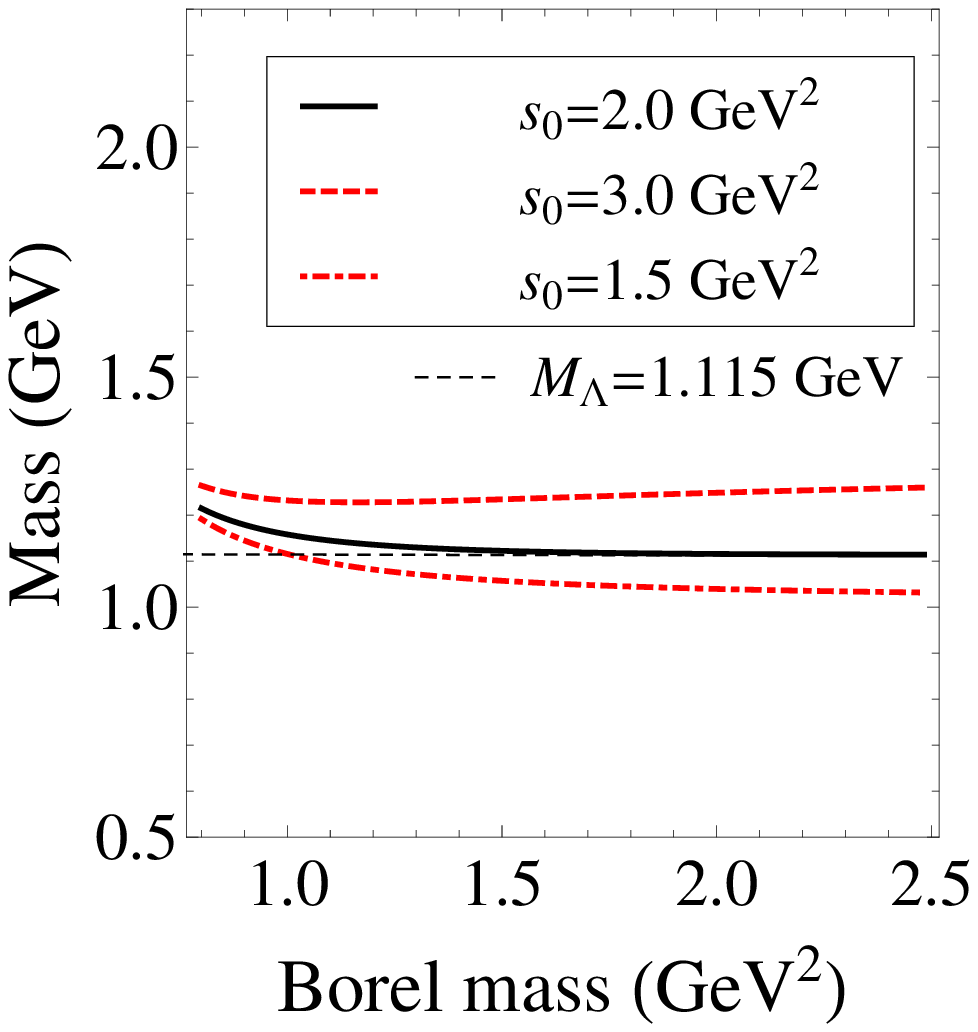}
\includegraphics[width=4.25cm,height=4.7cm]{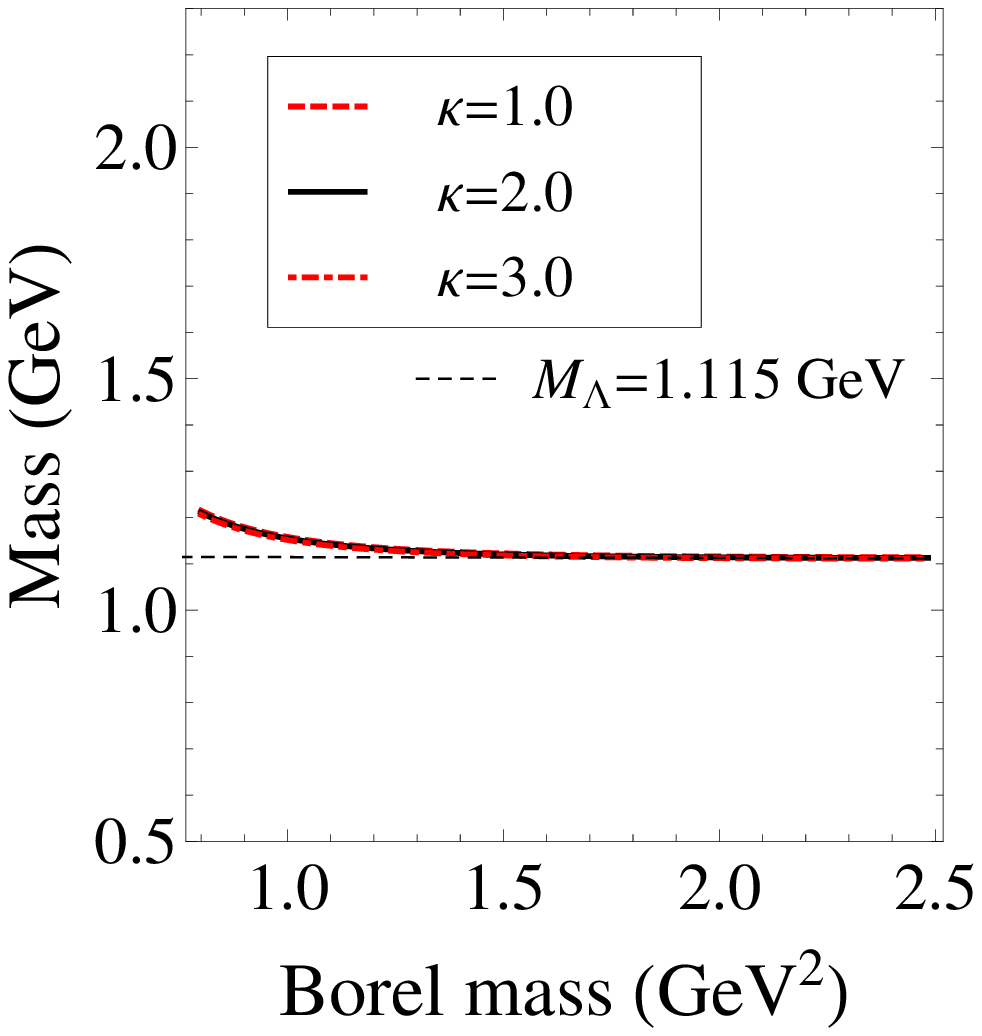}
\caption{(Color online) Borel curves for $\Lambda$ with different threshold  paramters (left) and with different $\kappa$ (right).}
\label{fig-lambda-s0}
\end{figure}

The Borel window in this case is set between $M^{2}=1.3$ 
and $M^{2}=2.2$ GeV$^{2}$. The lower limit is determined by the OPE
convergence where the contribution of the condensate part is less
than 30\% of the perturbative part, while the upper limit is determined by the
pole dominance in which the $\Lambda$ pole contribution exceeds the
contributions coming from the second term of Eq.~(\ref{pole}). Note
that the present sum rule has a much wider Borel window and better
stability than the usual baryonic QCD sum rule. The threshold $s_{0}$ was set at 2 GeV$^{2}$ as in a previous study~\cite{Narrison}.  The sum rule is rather stable under a small variation in the threshold parameter leading to the same diquark parameters. For larger variation, one finds  that the sum rule becomes less stable, as can be seen in the left figure of Fig.~\ref{fig-lambda-s0}.

Next we discuss diquark parameter dependence of the sum rule
stability. When $\lambda$ becomes smaller than $-0.5$ for
$m_{\phi}=0.2$ GeV, the Borel stability is lost and the mass
monotonically decreases toward smaller Borel mass.  For larger
$\lambda$ the curve breaks down at lower Borel mass.  Similar
behavior is found for a larger diquark mass of 0.3 GeV, as seen in
the right figure.  On the other hand, this time, the stable curve is
obtained when the diqurk condensate is smaller $\lambda=0.2$.   One
can further analyze the sum rule for different values of the diquark
mass and diquark condensate and find a strong correlation between the 
two parameters.
%that there is a duality type relation among the two parameters.
That is to say, to reproduce the physical $\Lambda$ mass with a
stable Borel curve, for larger (smaller)  diquark mass, we need
smaller (larger) diquark condensate. In general, obtaining a stable
curve should have enough information to determine a unique pair of
the parameters, since it is not trivial at all that the sum rule
works well with the Borel stability. Nevertheless, in the present
case, we find a duality-type relation among the two parameters. This
means that the $\Lambda$ mass may be essentially controlled by a
single physical quantity, such as a ``constituent" diquark mass. In
the limit when the diquark condensate is zero ($\lambda=0$), one
finds that the necessary diquark mass is around 0.4 GeV, which
corresponds to the constituent diquark mass.

Finally, we show the sum rule with $(\lambda,m_\phi)=(-0.5,02 {\rm GeV})$ but with different $\kappa$.  As can be seen in the right figure of Fig.~\ref{fig-lambda-s0}, the dashed and dot-dashed lines are almost indistinguishable from the solid line, suggesting a very weak dependence on the higher dimensional operators.

\subsection{$\Lambda_c$ and $\Lambda_b$}

\begin{figure}
\includegraphics[width=4.25cm,height=4.6cm]{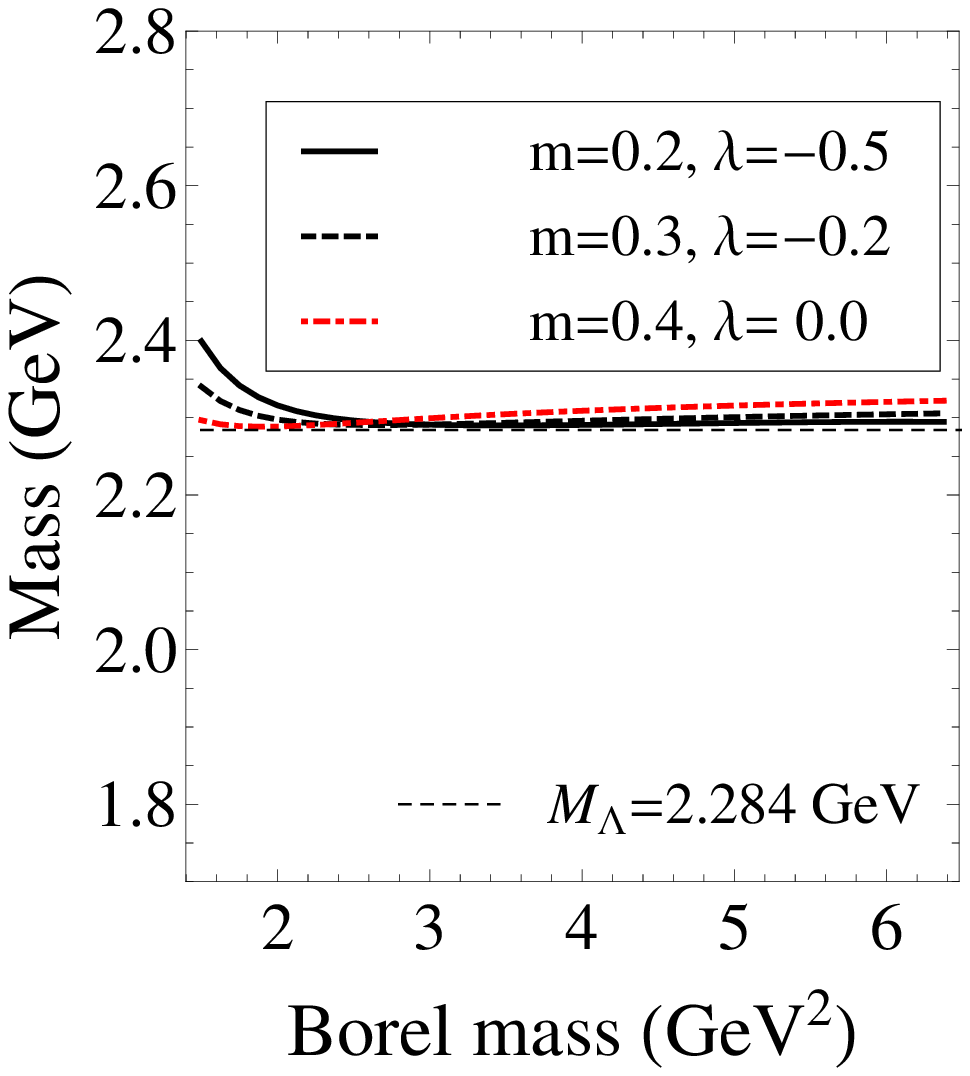}
\includegraphics[width=4.25cm,height=4.6cm]{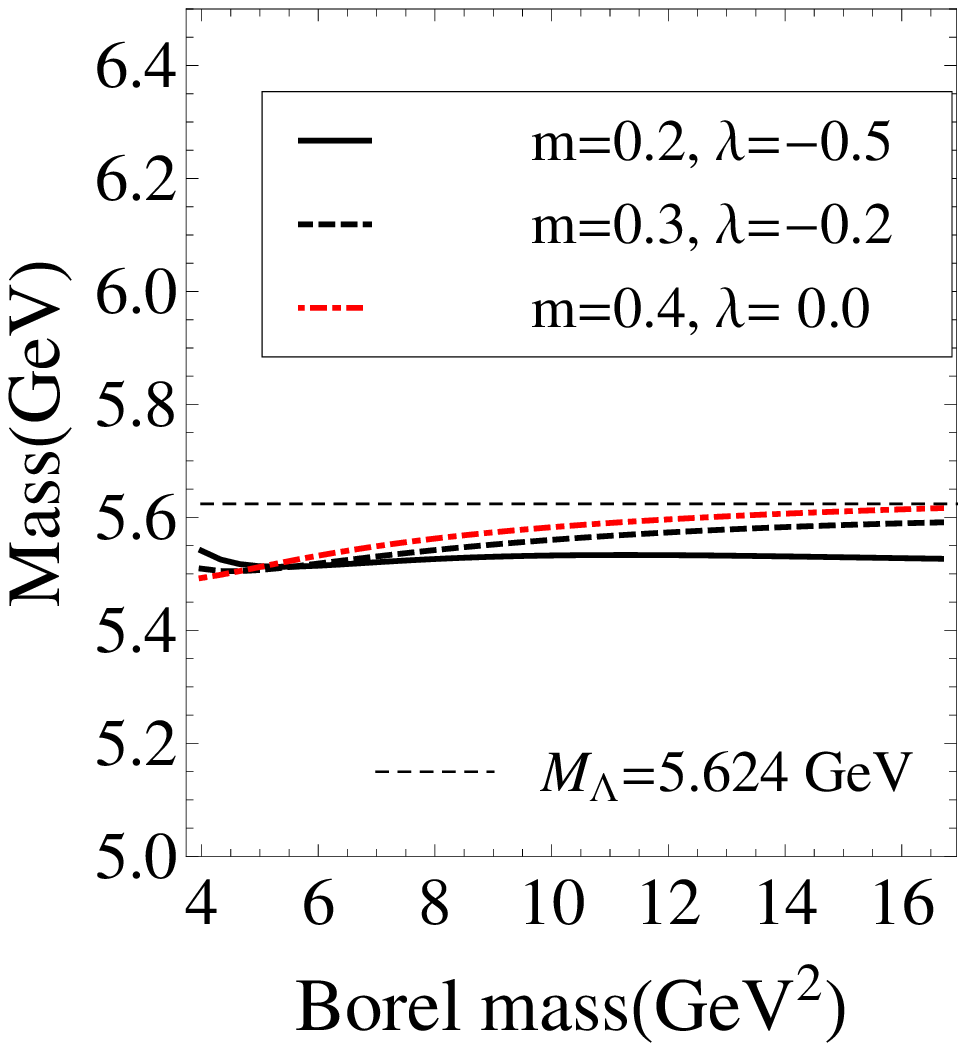}
%\Psfig{8.5cm}{lambdac.eps}
%~\Psfig{8.5cm}{LHC-mass.eps}
\caption{(Color online) Borel curves for $\Lambda_c$ (left) and $\Lambda_{b}$ (right) with the good diquark parameters.}
\label{fig-lambdac}
\end{figure}

We also calculate the $\Lambda_{c}$ and $\Lambda_{b}$ masses using
the good diquark parameters obtained in the $\Lambda$ sum rule, we
set $s_{0}=$7  and 40
GeV$^{2}$, respectively. Figure~\ref{fig-lambdac} shows the Borel
curves for $\Lambda_c$ and $\Lambda_b$ obtained with the sum rule
given in Eq.~(\ref{borel1}). As can be seen in the figure, all sets
of parameters $(\lambda, m_\phi)$ that worked well for $\Lambda$
reproduce the observed $\Lambda_c$ very well.  Although the best Borel curve slightly underestimates the mass of $\Lambda_b$, overall our results
imply that the diquark description works well again, suggesting
that, in these $\Lambda_h$ baryons, the diquark configuration represents important degrees of freedom having universal properties.
%%%
%the same set of parameters $(\lambda, m_\phi)$ that worked well for $\Lambda$ works well for $\Lambda_c$. Again the sum rule with the diquark description works well.
%This suggests that the diquark configuration is an important degrees of freedom in both $\Lambda$ and $\Lambda_{c}$ having a universal properties.

\begin{figure}[b]
\includegraphics[width=4.25cm,height=4.6cm]{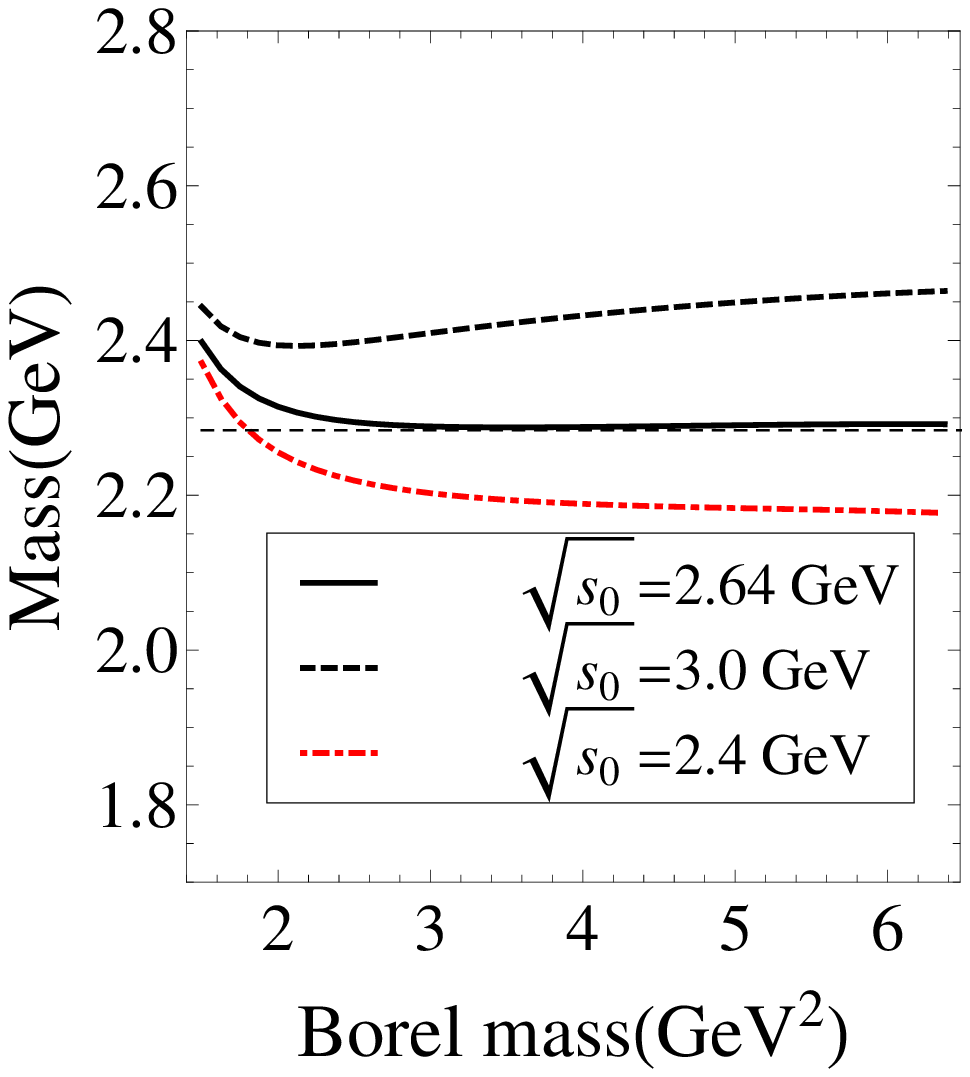}
\includegraphics[width=4.25cm,height=4.6cm]{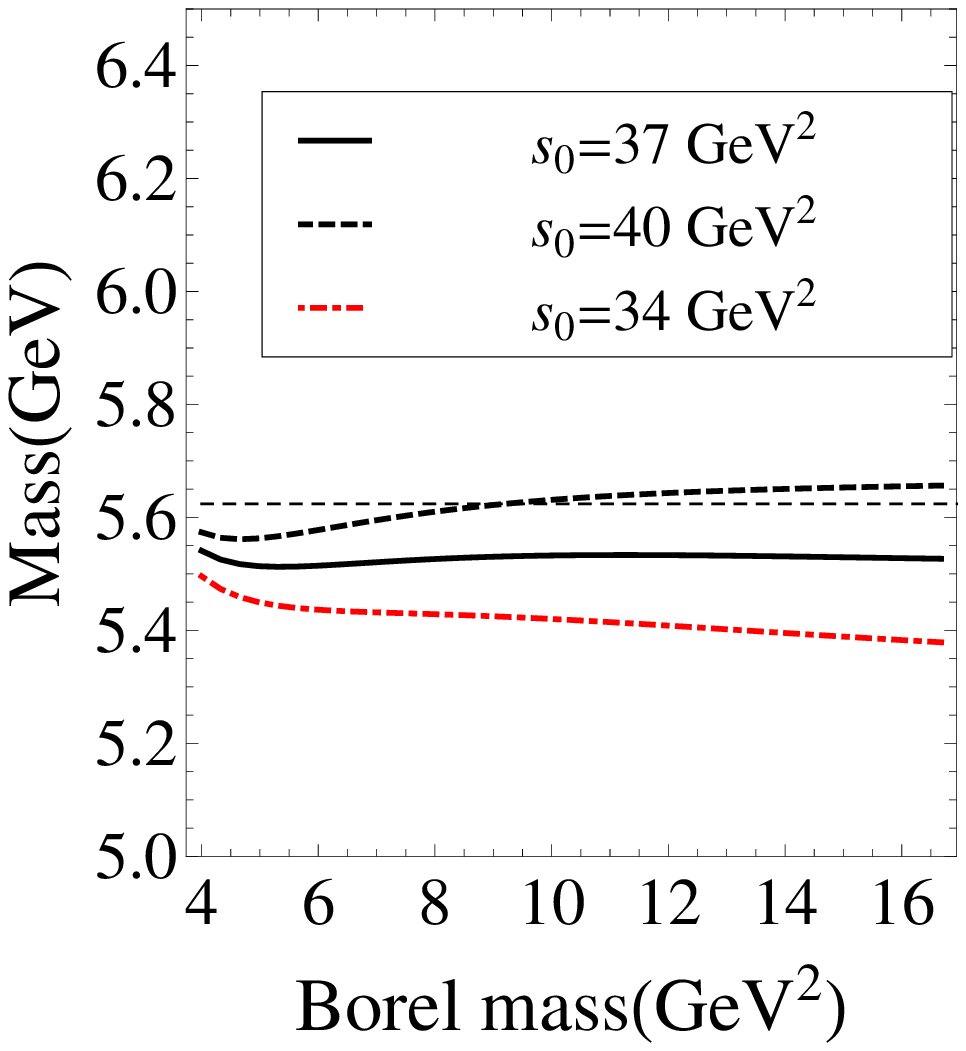}
\caption{(Color online) Borel curves showing the different choices of
$\sqrt{s_0}$ for $\Lambda_c$ (left) and $\Lambda_{b}$ (right) with the solid lines showing the case with the chosen thresholds. } \label{fig-lambdac-s0}
\end{figure}

In Fig.~\ref{fig-lambdac-s0}, we show the stability in the
choice of the continuum threshold by plotting the Borel curve for different $s_0$ values. Here, we plot the graphs for one set of the  accepted value of  $(\lambda,m_\phi)=(-0.5, 0.2 {\rm GeV})$. As can be seen from the figure, we
obtain a reasonable mass of $\Lambda_c$ when choosing the value of
$s_0$ which gives the most stable Borel curve.  Unfortunately, the threshold parameter that gives the most stable Borel curve slightly underestimates the mass of $\Lambda_b$.  We believe that the large difference in the bottom quark mass requires the use of different renormalization points for the vacuum expectation values, which can be a topic of future study.

\subsection{Summary and discussions}

The calculation of the correlation function as given in Eq.~(\ref{def})
means performing an OPE, assuming that the diquark structure is
intact even when the relative momentum between the diquark effective
field and the heavy quark is large.  This is a simplification, but
since we are interested only in the structure of the physical
$\Lambda$, all we need is that the approximation be valid for the
physical $\Lambda$, $\Lambda_c$ and $\Lambda_{b}$, as has been
shown through  our QCD sum rule analysis.

On the other hand, when we are analyzing states that have another
light quark or an antiquark, the diquark description is expected to
fail.  For example, in the proton, the $ud$ diquark, despite its
strong correlation, will tunnel into a new $du$ diquark with the
spectator $u$ quark.  In our language, we would need an effective
$\phi-u-d$ coupling to include this process.  The actual application of
our sum rule  to the nucleon case obtained by taking $m_s$ to zero
gives a nucleon mass above 1 GeV without a stable Borel curve. In that sense the minimal Lagrangian for the diquark given in Eq. (2) is unique for describing $\Lambda_h$.

We have also analyzed the sum rule with a $J_\sigma=\phi^\dagger
\phi$ current.  In the scalar nonet picture, the $\sigma$ meson is
composed of four quarks, but will have a strong quark-antiquark
component.  So we expect the coupling of our current with the
physical $\sigma$ to be very small.  Indeed, we find that the sum
rule obtained does not reproduce the $\sigma$ mass around 400 to 600
MeV.  Instead we find a monotonically decreasing Borel curve where
the mass ranges from 1400 to 1000 GeV.  The explicit treatment of four
quark fields in the QCD sum rule has reproduced a lower mass with a
stable curve~\cite{Kojo:2008hk}.
%\section{Summary}

In conclusion, we have introduced an explicit diquark degree of
freedom and analyzed the sum rule for $\Lambda$, $\Lambda_c$ and
$\Lambda_{b}$.  We find that we can reproduce the masses of these
states well, suggesting that these states can be represented by a
strong diquark and a spectator quark.  We further find that there is
a duality between diquark condensate and diquark mass from which we
can estimate the constituent diquark mass to be around 400 MeV.  We
expect that such an explicit diquark degree of freedom can be an
effective method to describe configurations where diquark
correlation is expected to remain strong.

%\begin{acknowledgments}
{\it Acknowledgments:} This work was supported by the Korean Ministry of
Education through the BK21 Program and Project No. KRF-2006-C00011. The work of
D.J.\ is supported by the Grant for Scientific Research
No.~22105507 and No.~22740161 from MEXT of Japan. A part of this
work was done in the Yukawa International Project for Quark-Hadron
Sciences (YIPQS).
%\end{acknowledgments}

%\bibliography{adsqcd,charm,chiral,chpt,eos,exotic,experiment,hydro,jpsi_sup,lattice,morita,nuclearmatter,pdg,qcd,qgp,QGPreview,qm08,sumrule,textbook,tft,thermalmodel,strange}

\end{document}